\documentclass[twocolumn, aps,prl,superscriptaddress, longbibliography]{revtex4-2}

\usepackage{graphicx}
\usepackage{dcolumn}
\usepackage{bm}
\usepackage{siunitx}

\usepackage{color}
\usepackage{booktabs}
\usepackage{miller}
\usepackage{wasysym} 
\usepackage{scalerel}
\usepackage{siunitx}
\usepackage{lineno}

\usepackage{hyperref}
\definecolor{r}{rgb}{0.86,0.08,0.23}
\definecolor{blue_n}{rgb}{0.,0.3,0.5}

\hypersetup{
backref=true,
pagebackref=true,
hyperindex=true,
colorlinks=true,
breaklinks=true,
citecolor = blue_n,
urlcolor= blue_n,
linkcolor= black,
pdftitle={A single optically detectable tumbling spin in silicon},
pdfauthor={F\'elix~Cache, Yoann Baron, Baptiste Lefaucher, Jean-Baptiste Jager, Frédéric Mazen, Frédéric Milési, Sébastien Kerdilès, Isabelle~Robert-Philip, Jean-Michel~G\'erard, Guillaume~Cassabois Vincent~Jacques and Ana\"is~Dr\'eau }
}

\begin{document}

    \title{A single optically detectable tumbling spin in silicon}

\author{F\'elix Cache}		\affiliation{Laboratoire Charles Coulomb, Universit\'e de Montpellier and CNRS, 34095 Montpellier, France}
\author{Yoann Baron} \affiliation{Université Grenoble Alpes, CEA-LETI, Grenoble 38000, France}
\author{Baptiste~Lefaucher}	\affiliation{Université Grenoble Alpes, CEA, Grenoble INP, IRIG, PHELIQS, 38000 Grenoble, France}
\author{Jean-Baptiste~Jager}	\affiliation{Université Grenoble Alpes, CEA, Grenoble INP, IRIG, PHELIQS, 38000 Grenoble, France}
\author{Frédéric Mazen} \affiliation{Université Grenoble Alpes, CEA-LETI, Grenoble 38000, France}
\author{Frédéric Milési} \affiliation{Université Grenoble Alpes, CEA-LETI, Grenoble 38000, France}
\author{Sébastien Kerdilès} \affiliation{Université Grenoble Alpes, CEA-LETI, Grenoble 38000, France}
\author{Isabelle~Robert-Philip}   \affiliation{Laboratoire Charles Coulomb, Universit\'e de Montpellier and CNRS, 34095 Montpellier, France}
\author{Jean-Michel~G\'erard}	\affiliation{Université Grenoble Alpes, CEA, Grenoble INP, IRIG, PHELIQS, 38000 Grenoble, France}
\author{Guillaume Cassabois}	 \affiliation{Laboratoire Charles Coulomb, Universit\'e de Montpellier and CNRS, 34095 Montpellier, France}
		\affiliation{Institut Universitaire de France, 75231 Paris, France.}
\author{Vincent Jacques}	 \affiliation{Laboratoire Charles Coulomb, Universit\'e de Montpellier and CNRS, 34095 Montpellier, France}
\author{Ana\"is Dr\'eau}	\email{anais.dreau@umontpellier.fr}	 \affiliation{Laboratoire Charles Coulomb, Universit\'e de Montpellier and CNRS, 34095 Montpellier, France} \email{anais.dreau@umontpellier.fr}

    \begin{abstract}

We demonstrate single spin spectroscopy of a fluorescent tumbling defect in silicon called the G center, behaving as a pseudo-molecule randomly reorienting itself in the crystalline matrix.  
Using high-resolution spin spectroscopy, we reveal a fine magnetic structure resulting from the spin principal axes jumping between discrete orientations in the crystal. 
Modeling the atomic reorientation of the defect shows that spin tumbling induces variations in the coupling to the microwave magnetic field, enabling position-dependent Rabi frequencies to be detected in coherent spin control experiments. 
By virtue of its pseudo-molecule configuration, the G center in silicon is a unique quantum system to investigate the mutual interaction between optical, spin and rotation properties in a highly versatile material.

    \end{abstract}

    \maketitle

Electron spin resonance spectroscopy is a widely used technique for analyzing the microscopic structure, local environment and reorientation of atomic and molecular systems~\cite{weil_electron_2006, schweiger_principles_2001,misra_molecular_2011}.
Conventional inductive detection methods typically require to probe more than a billion of electron spins such that single atom motion is hidden through ensemble averaging~\cite{weil_electron_2006}. 
While several single spin spectroscopy methods are currently available~\cite{wrachtrup_optical_1993, xiao_electrical_2004, elzerman_single-shot_2004, rugar_single_2004, komeda_observation_2011,wang_single-electron_2023}, they have been so far limited to static systems.   
Especially, spinning motion evidenced in ESR experiments~\cite{misra_molecular_2011}, such as top-like rotations~\cite{odonnell_origin_1983} or discontinuous-jump tumbling~\cite{brugh_spinning_2021}, has never been explored at the single-spin scale. 
In this letter, we optically detect the magnetic resonances of a single reorienting electron spin of an individual rotating color center in silicon, called the G center.

	\begin{figure}[h!]
		\includegraphics[width=\columnwidth]{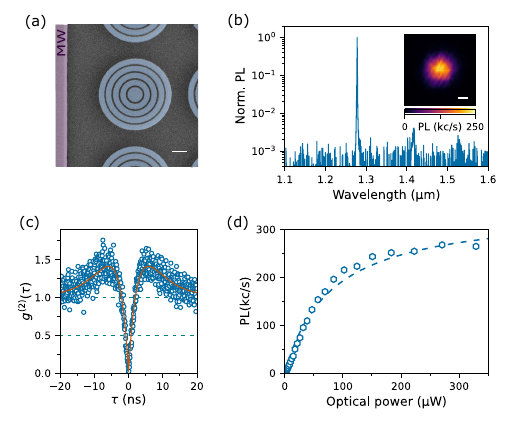}
		\caption{\textbf{A single G center integrated in a micro-cavity.}
(a) Colored scanning electron micrograph of a circular Bragg grating cavity positioned close to the microwave microstrip. 
Scale bar is 1 \unit{\micro\meter}.
(b) PL spectrum of an individual G center inside a cavity. 
Inset, PL map around the cavity. Scale bar is 1 \unit{\micro\meter}.
(c) Corresponding raw (uncorrected) autocorrelation function, giving $g^{(2)}(0) = 0.02 \pm 0.01$. 
Solid line is data fitting inferred from a rate equation model of a 3-level system~\cite{redjem_single_2020}. 
(d) Evolution of the PL count rate with optical pumping power. 
Dashed lines represent a fit to the data with a standard saturation function.}
		\label{fig:intro}
	\end{figure}

\begin{figure}[h!]
		\includegraphics[width=\columnwidth]{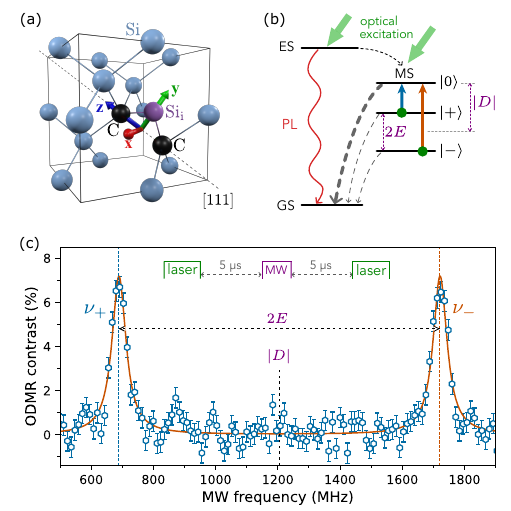}
		\caption{\textbf{ODMR on a single G center in silicon.}
(a) Atomic structure of the G center in silicon. 
(b) Energy level structure of the G center ~\cite{udvarhelyi_identification_2021}.
G center spin principal axes $\{\mathbf{x},\mathbf{y},\mathbf{z}\}$ are represented by the arrows (here $\{[\bar{1}10]$, $[\bar{1}\bar{1}2],[111]\}$~\cite{lee_optical_1982}).
(c) PL variation of a single G center versus MW excitation frequency. 
Laser and MW pulse durations: $1$ \unit{\micro\second} and $15$ ns, respectively. 
To assess the ODMR contrast, the defect PL signal integrated over the first 50 ns of the readout pulse is normalized with the signal when the MW is off-resonant.
 Solid line is data fitting with two Lorentzian functions. 
Error bars represent photon shot noise at one standard deviation (idem for all data plots). 
}
		\label{fig:odmr}
	\end{figure}

Known since the 70s, the G center has lately regained interest for its potential in the fast-growing field of silicon-based quantum technologies with individual color centers~\cite{redjem_single_2020, durand_broad_2021, simmons_scalable_2024, inc_distributed_2024, gritsch_optical_2025, lefaucher_bright_2025, dobinson_electrically_2025, song_entanglement_2026}.
The G center microscopic structure looks like a water molecule trapped in the silicon crystal~\cite{odonnell_origin_1983, udvarhelyi_identification_2021, ivanov_effect_2022}. 
There, the two hydrogen atoms are replaced by substitutional carbon atoms, and the middle oxygen atom by an interstitial silicon Si$_{\mathrm{i}}$ atom carrying two dangling bonds (Fig.~\ref{fig:odmr}(a)).
The reorientation of this pseudo-molecule is driven by the motion of the Si$_{\mathrm{i}}$ atom, that can occupy six different sites around the $\langle 111 \rangle$ axis~\cite{udvarhelyi_identification_2021}. 
The Si$_{\mathrm{i}}$ dynamics can range from perfectly delocalized states in unstrained $^{28}$Si bulk crystals to static positions in silicon-on-insulator (SOI) samples, where inherent strain reduces the G defect symmetry~\cite{durand_hopping_2024}. 
Single G centers offer two assets: firstly, single-photon emission at telecom wavelength~\cite{baron_single_2022, hollenbach_wafer-scale_2022, durand_genuine_2024} and secondly, a metastable electron spin triplet~\cite{lee_optical_1982, odonnell_origin_1983, vlasenko_spin-dependent_1986}. 
However, recent research has only dealt with harnessing its optical emission~\cite{lefaucher_cavity-enhanced_2023, prabhu_individually_2023,komza_indistinguishable_2024,  saggio_cavity-enhanced_2024, day_electrical_2024, ristori_strain_2024, kim_bright_2025,buzzi_spectral_2025, ma_nanoscale_2025}, letting its spin properties aside. 
Yet the G center was the first defect in silicon for which optical detection of magnetic resonance (ODMR) was reported~\cite{lee_optical_1982}. 
Since these pioneering ESR experiments performed on large ensembles in the 80s~\cite{lee_optical_1982, odonnell_origin_1983}, literature has been scarce on the spin properties of the G center in silicon.

In this paper, we investigate the reorientation of the electron spin of an individual G center integrated in a silicon photonic micro-cavity.
We perform high-resolution pulse ODMR spectroscopy to evidence a magnetic resonance fine structure, which results from spin tumbling between inequivalent crystal orientations. 
Transition-dependent Rabi frequencies are demonstrated and quantitatively interpreted by a model describing the single spin reorientation, providing information about the atomic arrangement of the G center atoms inside the silicon lattice.

A challenge for optical measurements on individual G centers in unstructured silicon is linked to their low photon count rates~\cite{hollenbach_wafer-scale_2022, durand_genuine_2024}, resulting from a small fraction of the emission being collected ($\sim 1\%$)~\cite{redjem_single_2020, durand_genuine_2024} and a weak radiative rate~\cite{komza_indistinguishable_2024}. 
To accelerate the emission rate and increase the collection efficiency~\cite{lefaucher_cavity-enhanced_2023, saggio_cavity-enhanced_2024, kim_bright_2025}, single G centers are integrated inside circular Bragg grating cavities fabricated in a SOI sample (Fig. \ref{fig:intro}(a), \cite{SI})~\cite{davanco_circular_2011, lefaucher_bright_2025, ma_nanoscale_2025}.
Experiments are performed below 10K in a low-temperature home-made confocal microscope described in \cite{SI}. 
The photoluminescence (PL) spectrum for a G defect inside a cavity shows a zero-phonon line at 1278 nm whose amplitude is 3 orders of magnitude above the noise level (Fig.~\ref{fig:intro}(b)). 
Effective Debye-Waller factor reaches 94\% here, compared to $\sim$18\% for G centers in unstructured SOI~\cite{beaufils_optical_2018}. 
To ensure that the cavity contains only one defect, we measure the autocorrelation function $g^{(2)}(\tau)$ of the PL signal under continuous excitation (Fig.~\ref{fig:intro}(c)) (\cite{SI}). 
At zero delay $\tau = 0$, the antibunching behavior is nearly perfect, as witnessed by an uncorrected $g^{(2)}(0) = 0.02 \pm 0.01$. 
This high-purity single-photon emission confirms the presence of a single emitter in the cavity. 
At intermediate delays, the photon bunching ($g^{(2)}(\tau)> 1$) indicates a non-radiative relaxation channel involving a long-lived metastable level~\cite{beveratos_room_2002}, likely the G center electron spin triplet~\cite{lee_optical_1982}. 
At high optical excitation power, the maximum PL signal recorded for this individual G defect is around 260 kcounts.s$^{-1}$ (Fig.~\ref{fig:intro}(d)). 
The zero-phonon line single-photon rate  is thus boosted by $\times 100$ with the cavity integration compared to isolated G centers in unpatterned SOI~\cite{hollenbach_wafer-scale_2022, durand_genuine_2024}.

To enable ODMR, the G center energy level structure must exhibit spin-dependent photo-dynamics~\cite{lee_optical_1982}. 
After non-resonant above-bandgap optical excitation, the defect has two possible relaxation pathways to the ground state (GS) spin singlet: either via the excited state (ES) spin singlet and radiative recombination, or through the metastable state (MS) spin triplet by non-radiative decay (Fig.~\ref{fig:odmr}(b))~\cite{lee_optical_1982}.
The MS spin triplet is governed by the zero-field splitting (ZFS) interaction induced from spin-spin interaction between its two unpaired electrons:
\begin{equation}
\hat{H}_0 = D \hat{S}_z^2 + E(\hat{S}_x^2-\hat{S}_y^2), 
\end{equation}
with $D<0$ and $E>0$ (sign from \cite{udvarhelyi_identification_2021}) being the longitudinal and transverse ZFS parameters~\cite{weil_electron_2006}. 
The spin principal axes set by the ZFS tensor are \{$\mathbf{x}: [\bar{1}10]$,  $\mathbf{y}:[\bar{1}\bar{1}2]$, $\mathbf{z}:[111]$\}, with $\mathbf{z}$ being aligned along the C-C direction (Fig.~\ref{fig:odmr}(a)). 
At zero magnetic field, the three spin eigenstates are therefore $|0\rangle \!=\! |m_s\! = \!0\rangle, |+\rangle, |- \rangle$, with $|\pm\rangle \!= \!{ (|m_s \! = \!+1\rangle }\pm {|m_s\! =\! -1 \rangle})/\sqrt{2}$. 
As they are not coupled at first order to the GS~\cite{udvarhelyi_identification_2021}, $|\pm\rangle$ states should have longer lifetimes than $|0\rangle$.
Consequently, optical pumping should trap the spin population into the two long-lived states $|\pm\rangle$. 
Then, when the frequency of a microwave (MW) magnetic field is at resonance with one of the two spin transitions at $\nu_{\pm}= |D\pm E|$, the spin is flipped to the short-lived state $|0\rangle$, leading to an increase of the G defect PL signal.

The ESR frequencies of the G center are probed using a sequence alternating laser pulses for spin preparation and readout, and MW pulses for spin rotation. 
As shown in Figure~\ref{fig:odmr}c, the pulse ODMR spectrum of a single G defect reveals two resonance frequencies at $\nu_+ = 689 \pm 4$ MHz and $\nu_- = 1721 \pm 4$~MHz (SI). 
Hence, we infer the ZFS parameters of the G~center in silicon: $|D|= 1205 \pm 6$ MHz and $E = 516 \pm 6$~MHz, respectively. 
The large transverse $E$ component is intrinsic to the $C_{1\mathrm{h}}$ low-symmetry of the G defect and indicates a strong deviation from axial symmetry~\cite{odonnell_origin_1983}. 
The ZFS values from this first ODMR spectrum reported on a single G center are in perfect agreement with previous ODMR measurements performed on dense G ensembles~\cite{lee_optical_1982, odonnell_origin_1983, vlasenko_spin-dependent_1986}. 
ODMR on an individual G defect is key to unveiling the fine magnetic structure resulting from spin tumbling, which is hidden in ensemble measurements. 
However, at this stage, there is no evidence of spin reorientation since the ESR lines are broadened by the strong MW magnetic field driving the spin transitions~\cite{dreau_avoiding_2011}.

	\begin{figure}[h!]
		\includegraphics[width=\columnwidth]{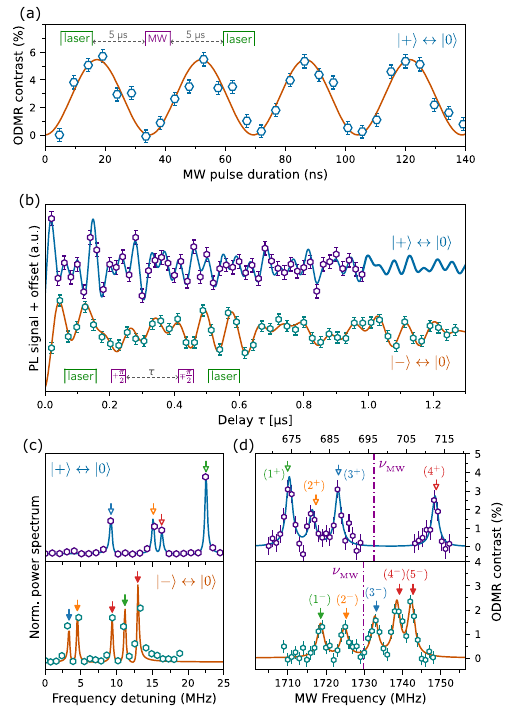}
		\caption{\textbf{Spin coherence properties and fine spin structure of a single G center.}
(a) Rabi oscillations on ${|+\rangle \leftrightarrow |0\rangle}$, fitted with a sine function (solid line). 
(b) Ramsey fringes measured on the 2 spin transitions with MW frequency $\nu_{_{\mathrm{MW}}}$. 
Solid lines represent data fitting with a sum of 4 (\textit{top}) and 5 (\textit{bottom}) cosine functions with an exponential decay envelope, providing $T_{2_{\nu_+}}^* = 0.8 \pm 0.2$ \unit{\micro\second} and  $T_{2_{\nu_-}}^* = 1.1 \pm 0.3$ \unit{\micro\second}.
(c) Corresponding Fourier-transform power spectra of the data (markers) and of the fit results (solid lines). 
(d) High-resolution ODMR spectra recorded with the same sequence as in Figure~\ref{fig:odmr}(c) but with a MW $\pi$-pulse duration about 300 ns. 
Solid curves are fits with Lorentzian functions. 
Color arrows show the frequencies extracted from data fitting of Ramsey signals. }
		\label{fig:rabi}
	\end{figure}

To get rid of power broadening, one way is to rely on Ramsey spectroscopy, where spectral resolution is limited by the coherence time $T_2^*$~\cite{schweiger_principles_2001}. 	
To do so, we first perform Rabi oscillations to adjust the correct MW pulse durations. 
Figure~\ref{fig:rabi}(a) shows that the ODMR contrast oscillates with the MW pulse duration, with a typical $\pi/2$ pulse of $\simeq$ 9 ns. 
The data of the Ramsey sequence applied to the two electron spin transitions are plotted in Figure~\ref{fig:rabi}(b). 
The coherent signal envelope follows a mono-exponential decay with time constant $T_{2_{\nu_+}}^* = 0.8 \pm 0.1$ \unit{\micro\second} and  $T_{2_{\nu_-}}^* = 1.1 \pm 0.2 $ \unit{\micro\second}. 
Most importantly, the Ramsey signals display multiple-frequency beatings indicating that several spin transitions with different detuning with the MW frequency are involved. 
Indeed, data fitting and associated Fourier-transform spectra show 4 and 5 frequency peaks for the $|+\rangle \leftrightarrow |0\rangle$ and $|-\rangle \leftrightarrow |0\rangle$ spin transitions, respectively (Fig. \ref{fig:rabi}(c)). 
This fine structure, previously hidden in the power-broadened ODMR spectrum from Figure~\ref{fig:odmr}(c), is confirmed by performing high-resolution ODMR spectroscopy.
As shown in Figure~\ref{fig:rabi}(d), all the fine spin transitions are directly resolved in ODMR spectra using the same pulse sequence as in Figure~\ref{fig:odmr}(c), but with MW $\pi$ pulses whose duration approaches $T_{2_{\nu_\pm}}^*$~\cite{dreau_avoiding_2011}. 
In these high-resolution ODMR spectra, this fine spin structure cannot be explained by hyperfine interaction with nuclear spins ($^{13}$C and $^{29}$Si, $I = 1/2$)~\cite{weil_electron_2006}. 
Instead, this ODMR line pattern is the signature of the single electron spin tumbling of the G center inside the silicon crystal.

\begin{figure}[h!]
		\includegraphics[width=\columnwidth]{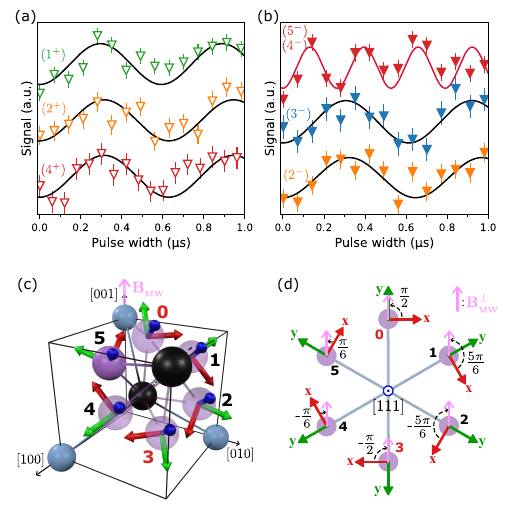}
		\caption{\textbf{A reorienting spin.}
(a-b) Frequency selective Rabi oscillations, recorded with MW frequency at resonance, except for the $(4^-,5^-)$ lines that partially overlap and for which the MW frequency was set at the center frequency. 
Once corrected from a 1.9-MHz detuning~\cite{cohen-tannoudji_quantum_2019}, the Rabi frequency associated to the $(4^-,5^-)$ lines is $\Omega = 3.4 \pm 0.1$ MHz. 
For the other lines, the Rabi frequency is $\Omega' = 1.6 \pm 0.1$ MHz.  
(c) Microscopic structure of the G center showing \textbf{x} and \textbf{y} spin axes (red and green arrows, resp.) for the 6 possible sites of the Si$_{\mathrm{i}}$ atom (purple) and $\mathbf{B}_{\scaleto{\mathrm{MW}}{4pt}}$ aligned along $[001]$ (\cite{SI}).
(d) Scheme representing, for each Si$_{\mathrm{i}}$ position,  the azimuthal angle $\varphi$ value and $\textbf{x}, \textbf{y}$ and $\mathbf{B}^{\perp}_{\scaleto{\mathrm{MW}}{4pt}}$ (pink arrow) in the $(111)$-plan.
}
		\label{fig:reorient}
	\end{figure}

The G center spin reorientation is evidenced by recording frequency-selective Rabi oscillations recorded with long MW $\pi$ pulses. 
For the low-frequency ESR branch $|+\rangle \leftrightarrow |0\rangle$, at a given MW power, all fine spin transitions display the same Rabi frequency (Fig.~\ref{fig:reorient}(a)). 
On the contrary, for the other ESR branch, the Rabi oscillations with excitation selective on the two fine spin transitions $(4^-,5^-)$ are roughly twice faster than for the other transitions, as shown in Figure~\ref{fig:reorient}(b). 
The extracted ratio of Rabi frequencies at resonance is ${\Omega/\Omega'=2.2\pm0.2}$.
The MW magnetic field can be written as ${\mathbf{B}_{\scaleto{\mathrm{MW}}{4pt}} \cos(2\pi\nu_{\scaleto{\mathrm{MW}}{4pt}} t)}$.
When $\nu_{\scaleto{\mathrm{MW}}{4pt}}$ is resonant with the spin transitions, the Rabi frequencies in the rotating wave approximation express as~\cite{cohen-tannoudji_quantum_2019}: 
\begin{equation}
\Omega_{\pm} = |\langle 0| -\gamma_e \mathbf{B}_{\scaleto{\mathrm{MW}}{4pt}} \cdot \hat{\mathbf{S}} | \pm \rangle| , 
\label{eq:rabi0}
\end{equation}
with $\gamma_e$ being the electron spin gyromagnetic ratio~\cite{lee_optical_1982, odonnell_origin_1983}. 
Since the orientation of $\mathbf{B}_{\scaleto{\mathrm{MW}}{4pt}}$ is fixed (\cite{SI}), observing different Rabi frequencies implies that the principal axes of the spin operator $\hat{\mathbf{S}}$ do rotate over time.

The electron spin tumbling of the G center can be modeled by considering the motion of its Si$_{\mathrm{i}}$ atom and the spin selection rules resulting from the spin mixing induced by the transverse ZFS component~\cite{kolbl_determination_2019}. 
The spin operator coupling the coherent superposition $|+\rangle$ to the state $|0\rangle$ is $\hat{S}_x$ and, conversely, $\hat{S}_y$ couples the two states ${|-\rangle  \leftrightarrow |0\rangle}$. 
Following Eq.~(\ref{eq:rabi0}), the MW magnetic field components driving these two spin transitions are thus $B_x$ and $B_y$, respectively. 
By defining $B_{\scaleto{\mathrm{MW}}{4pt}}^{\perp}= \sqrt{{\mathrm{B}_{\scaleto{\mathrm{MW}}{4pt}}^{(x)}}^2+{\mathrm{B}_{\scaleto{\mathrm{MW}}{4pt}}^{(y)}}^2}$ and $\varphi$ the azimuthal angle of $\mathbf{B}_{\scaleto{\mathrm{MW}}{4pt}}$ in the spin coordinate system, the Rabi frequency expressions can be simplified to:
\begin{equation}\label{eq:rabix}
\Omega_+  =  |\gamma_e B_{\scaleto{\mathrm{MW}}{4pt}}^{(x)}| =| \gamma_e B_{\scaleto{\mathrm{MW}}{4pt}}^{\perp}| \cdot|\cos \varphi| ,
\end{equation}
\begin{equation}\label{eq:rabiy}
\Omega_-  =  |\gamma_e B_{\scaleto{\mathrm{MW}}{4pt}}^{(y)}| = |\gamma_e B_{\scaleto{\mathrm{MW}}{4pt}}^{\perp}| \cdot|\sin \varphi| .
\end{equation}
The Si$_{\mathrm{i}}$ can sit on one of six possible positions $\{\mathbf{0}-\mathbf{5}\}$ sharing the same \textbf{z} axis but with different \textbf{x} and \textbf{y} directions, as shown in Figures \ref{fig:reorient}(c-d)~\cite{odonnell_origin_1983, udvarhelyi_identification_2021}. 
$\mathbf{B}_{\scaleto{\mathrm{MW}}{4pt}}$ is aligned along $[001]$, set by the sample geometry (\cite{SI}).
Based on the azimuthal angle $\varphi$, we can distinguish two situations, depending on whether the G orientation corresponds to $\{\mathbf{0},\mathbf{3}\}$ or to $\{\mathbf{1},\mathbf{2},\mathbf{4},\mathbf{5}\}$.
When the interstitial silicon atom is either in site \textbf{0} or \textbf{3}, $\mathbf{B}_{\scaleto{\mathrm{MW}}{4pt}}$ is perpendicular to the \textbf{x} spin axis (i.e. $|\varphi| = \pi/2$), so the spin transition ${|+\rangle \leftrightarrow |0\rangle}$ cannot be driven by the MW magnetic field generated by the microstrip. 
This is not the case for the 4 other Si$_{\mathrm{i}}$ positions $\{\mathbf{1},\mathbf{2},\mathbf{4},\mathbf{5}\}$, for which $ |\varphi| = \pi/6$ or $5\pi/6$ (Fig. \ref{fig:reorient}(d)). 
As a consequence, there are a maximum of 4 fine spin transitions observable for the lower ESR branch ${|+\rangle \leftrightarrow |0\rangle}$, in agreement with the experimental data in Figure~\ref{fig:rabi}(b-c). 
For the other ESR branch, the spin  ${|-\rangle \leftrightarrow |0\rangle}$ is allowed for all the six G defect orientations as $|\mathrm{B}_{\scaleto{\mathrm{MW}}{4pt}}^{(y)}| \neq 0$ (Fig. \ref{fig:reorient}(a)).
In the fine structure from Figure \ref{fig:rabi}(d), we only resolve five spin resonance lines, possibly because the spin transitions of two sites are at similar frequencies. 

The spin tumbling model describes as well the behavior of the Rabi oscillations.  
First, it predicts that all G spin orientations within a subset share the same Rabi frequency for a given ESR transition (Eq.~\ref{eq:rabix}-\ref{eq:rabiy}).
As displayed in Figure~\ref{fig:reorient}(a), all coherent oscillations performed on the fine transitions of ${|+\rangle \leftrightarrow |0\rangle}$ evolve in phase. 
For the electron spin transition ${|-\rangle \leftrightarrow |0\rangle}$, the model indicates that the ratio of the Rabi frequency between the two sets $\{\mathbf{0},\mathbf{3}\}$ and $\{\mathbf{1},\mathbf{2},\mathbf{4},\mathbf{5}\}$ is directly equal to $ \Omega_-^{(0)}/\Omega_-^{(1)}=\sin(\pi/2)/\sin (\pi/6)=2$. 
This estimation is in very good agreement with the experimental results on ${|-\rangle \leftrightarrow |0\rangle}$, showing two spin transitions with a Rabi frequency $\Omega/\Omega'=2.2\pm 0.2$ higher than the other fine transitions (Fig.~\ref{fig:reorient}(b)).
This model explains also the different shapes of Rabi oscillations recorded at high MW power (\cite{SI}).

This single spin tumbling effect can be used to extract information on the microscopic structure of the G center in silicon in its metastable level. 
First of all, the two fine spin transitions linked to larger Rabi frequency, $(4^-,5^-)$, correspond to the G center being in either position $\{\mathbf{0}\}$ or $\{\mathbf{3}\}$. 
The rest of the fine spin transitions are associated to the other set of defect orientations: $\{\mathbf{1},\mathbf{2},\mathbf{4},\mathbf{5}\}$. 
Observing this fine spin structure implies that the ZFS interaction is different for each of the six defect orientations. 
In comparison, optical spectroscopy on individual G centers in SOI samples has so far only revealed a partial lift of the degeneracy of the electric dipolar transitions~\cite{durand_hopping_2024}. 
The origin of this extrinsic perturbation that varies from one G center to another (\cite{SI}) could be lattice strain resulting from the SOI stack~\cite{fukuda_white_2006} or inhomogeneous electric field generated by surrounding charges~\cite{dolde_nanoscale_2014}.

In conclusion, a single electron spin tumbling mechanism inside a silicon crystal is unveiled using ODMR spectroscopy on individual G centers integrated in photonic cavities.
The spin jumping over time between inequivalent crystal orientations leads to a fine spin structure with different coupling to the MW magnetic field. 
This effect is evidenced in high-resolution spin spectroscopy and frequency-selective coherent spin control experiments, and quantitatively explained by a model based on simple geometric arguments that describes the single spin reorientation.

In a near future, spin spectroscopy could be used to probe the control over the G center reorientation dynamics, including freezing the spin tumbling effect under optical resonant excitation~\cite{durand_hopping_2024}, temperature-activated top-like rotation~\cite{lee_optical_1982, odonnell_origin_1983, vlasenko_spin-dependent_1986} or coherent delocalized rotational states for perfectly symmetric defects~\cite{udvarhelyi_identification_2021}.
The physical complexity of the G center could be further enriched with hyperfine interaction with nuclear spins ($^{29}$Si or $^{13}$C).
Due to its unusual pseudo-molecule geometry, the G center in silicon offers a rich physical system with interplaying telecom photon, spin and rotation degrees of freedom in the most mature platform for nanofabrication.

\section*{Acknowledgments}
This work is supported by  the Plan France 2030 through the project OQuLuS (No.~ANR-22-PETQ-0013), the French National Research Agency (ANR) through the project WOUAH (No.~ANR-24-CE47-4667), CEA through the PTC-MP “W-TeQ”  internal project, and the European Research Council (ERC) under the European Union’s  Horizon 2020 research and innovation programme (project SILEQS, Grant No.~101042075).  B.L. is supported by the “Program QuantForm-UGA No. ANR-21-CMAQ-003 France 2030” and “Laboratoire d’Excellence LANEF No. ANR-10-LABX-51-01”. The authors acknowledge the assistance of the staff of the Grenoble Advanced Technological  Platform (PTA) and of the CEA Leti cleanroom facility.

	\bibliography{arXiv_V2_single_spin_tumbling}

	\end{document}